# Notes on Recent Approaches Concerning the Kirchhoff-Law-Johnson-Noise-based Secure Key Exchange

Laszlo B. Kish [a], Tamas Horvath [b,c]

[a] Department of Electrical and Computer Engineering, Texas A&M University, College Station, TX 77843-3128, USA; email: Laszlo.Kish@ece.tamu.edu

[b] Department of Computer Science, University of Bonn, Germany
[c] Fraunhofer IAIS, Schloss Birlinghoven, D-53754 Sankt Augustin, Germany; email: tamas.horvath@iais.fraunhofer.de

**Abstract.** We critically analyze the results and claims in [Physics Letters A 373 (2009) 901–904].

We show that the strong security leak appeared in the simulations is only an artifact and not caused by "multiple reflections". Since no wave modes exist at cable length of 5% of the shortest wavelength of the signal, no wave is present to reflect it.

In the high wave impedance limit, the conditions used in the simulations are heavily unphysical (requiring cable diameters up to 28000 times greater than the measured size of the known universe) and the results are modeling artifacts due to the unphysical values.

At the low cable impedance limit, the observed artifacts are due to violating the recommended (and tested) conditions by neglecting the cable capacitance restrictions and using about 100 times longer cable than recommended without cable capacitance compensation arrangement.

We implement and analyze the general circuitry of Liu's circulator [Physics Letters A 373 (2009) 901–904] and confirm that they are conceptually secure against passive attacks. We introduce an asymmetric, more robust version without feedback loop. Then we crack all these systems by an active attack: a circulator-based man-in-the middle attack.

Finally, we analyze the proposed method to increase security by dropping only high-risk bits. We point out the differences between different types of high-risk bits and show the shortage of this strategy for some simple key exchange protocols.

## 1. Introduction

In a recent paper [1] a circulator-based development and computer simulations have been published about the secure key exchange system utilizing Kirchhoff's Law and Johnson-like noise (KLJN cypher) [2-6], which is, at the moment, the only classical physical system offering unconditional security in the idealized limit. If the claims in [1] were valid, they could be of high impact on the future of the KLJN type systems. In the present paper, we give a critical analysis of the results and claims in [1], especially about the circulator arrangement, the theory, the computer simulations of the original system proposed in [2] and their interpretations, and about the proposal of dropping high-risk bits to enhance security.

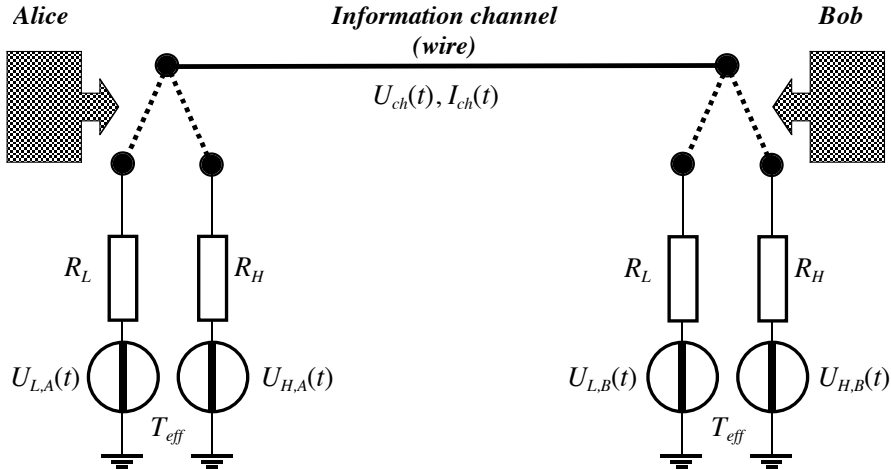

**Figure 1.** Model of the idealized KJLN cypher (secure key exchange system) [2].

First we briefly survey the necessary background to help the reader to follow the argumentations. Figure 1 shows the model of the idealized KJLN cypher designed for secure key exchange in [2]. The resistor $R_L$ and $R_H$ represent the low, $L$ (0), and high, $H$ (1), bits, respectively. At each clock period, Alice and Bob randomly choose one of the resistors and connect it to the wire line. The situation *LH* or *HL* represents secure bit exchange [2], while *LL* and *HH* are insecure. The Gaussian voltage noise generators (white noise with publicly agreed bandwidth) represent a corresponding thermal noise at publicly agreed effective temperatures $T_{eff}$ (typically $T_{eff} > 1$ billion Kelvins [3]). The power density spectra $S_{u,L}(f)$ and $S_{u,H}(f)$ of the voltages $U_{L,A}(t)$ and $U_{L,B}(t)$ supplied by the voltage generators in $R_L$ and $R_H$ are given by:

$$S_{u,L}(f) = 4kT_{eff}R_L \quad \text{and} \quad S_{u,H}(f) = 4kT_{eff}R_H \quad , \tag{1}$$

respectively.

In the case of secure bit exchange (*LH* or *HL* situations), the power density spectrum of channel voltage $U_{ch}(t)$ and channel current $I_{ch}(t)$ are given as (see [2] for further details):

$$S_{u,ch}(f) = 4kT_{eff} \frac{R_L R_H}{R_L + R_H} \tag{2}$$

and

$$S_{i,ch}(t) = \frac{4kT_{eff}}{R_L + R_H} \tag{3}$$

Observe that during the *LH* or *HL* situation, due to linear superposition, Equation-2 is the sum of the spectrum of two situations: When only the generator in $R_L$ is running:

$$S_{L,u,ch}(f) = 4kT_{eff}R_L \left(\frac{R_H}{R_L + R_H}\right)^2 \tag{4}$$

and when the generator in $R_H$ is running:

$$S_{H,u,ch}(f) = 4kT_{eff}R_H \left(\frac{R_L}{R_L + R_H}\right)^2 \tag{5}$$

The ultimate security of the system against passive attacks is provided by the fact that the power $P_{H \to L}$ with which resistor $R_H$ is heating $R_L$ is equal to the power $P_{L \to H}$ with which resistor $R_L$ is heating $R_H$ [2]. This follows directly from the second law of thermodynamics and the energy conservation law; the proof can also be derived from Equation 3 for a frequency bandwidth of $\Delta f$ by:

$$P_{L \to H} = \frac{S_{L,u,ch}(f)\Delta f}{R_H} = 4kT_{eff}\frac{R_L R_H}{(R_L + R_H)^2} \qquad (6a)$$

$$P_{H \to L} = \frac{S_{H,u,ch}(f)\Delta f}{R_L} = 4kT_{eff}\frac{R_L R_H}{(R_L + R_H)^2} \qquad (6b)$$

The equality $P_{H \to L} = P_{L \to H}$ (see Equations 6) is in accordance with the *second law of thermodynamics*; violating this equality would mean not only violating basic laws of physics, but also that the eavesdropper (Eve) could utilize the voltage-current crosscorrelation $\langle U_{ch}(t)I_{ch}(t) \rangle$, which is normally zero, to extract the bit [2].

Any deviations from this circuitry, including parasitic elements, inaccuracies, non-Gaussianity of the noise, etc. will cause potential information leak toward Eve [7-10]. For further analysis design aspects and comparisons, see [3,8,10]. The situation is similar to quantum key exchange systems, which are the only unconditionally secure competitors of the KLJN system. However, they assume idealistic situations (e.g. single photon source, noise-free channel) which are unreachable in a real physical system. The experimental results in [3] clearly demonstrate that the KLJN cypher outperforms the raw-bit security of conceptual quantum communicators. Because of the non-zero information leak due to non-ideality effects, both the KLJN and quantum communicators require privacy amplification [11] of the raw bits, which is a proper algorithmic transformation making a shorter key of high security from the longer raw key of low security.

To provide unconditional security against invasive attacks, the fully armed KLJN cypher system is monitoring the instantaneous current and voltage values, at both ends (i.e., Alice and also Bob) [3,12], and these values are compared either via broadcasting them or via an authenticated public channel. It is important to note that these current and voltage data contain all the information Eve can have. This implies that Alice and Bob have full knowledge about the information Eve may have; a particularly important property of the KLJN system that can be utilized in secure key exchange. This situation implies the following important features of the KLJN system:

**1.** The KLJN system is *always unconditionally secure*, even when it is non-ideal, in the following sense. The current and voltage data inform Alice and Bob about the exact information leak. Hence, they can always decide either to shot down the communication or to take the risk. Notice that this is a stronger property than that of quantum communicators.

**2.** Even when the communication is jammed by invasive attacks or inherent non-idealities in the KLJN system, the system remains secure because no information can be eavesdropped by Eve without the full knowledge of Alice and Bob about this potential incidence, and without the full information Eve might have extracted, see also Section 3.

**3.** The KLJN system is naturally and fully protected against the man-in-the-middle attack [12], *even during the very first run of the operation* when no hidden signatures can be applied yet. This feature is provided by the unique property of the KLJN system that zero bit information can only be extracted during a man-in-the-middle attack because the alarm goes on before the exchange of a single key bit has taken place [12].

The outline of the prototype of he KLJN cypher [3] is shown in Figure 2. The various non-idealities have been addressed by different tools with the aim that the information leak due to non-idealities should stay below 1% of the exchanged raw key bits. It is important to note that while 1% is about the lowest limit an idealistic quantum communicator may achieve, for the KLJN cypher it was 0.19% for the best attack [3]. In the present paper, we are only concerned about two of these tools (see also Section 2):

**(i)** The role of the line filter (and that of the band limitation of the noise generator) is to provide the no-wave limit in the cable (see Figure 2). That is, the shortest wavelength component in the driving noises should be much longer than the double of the cable length in order to guarantee that no active wave modes and related effects (e.g., reflection, invasive attacks at high frequencies, etc.) take place in the cable. While in [2] an upper limit for the cable length of 10% of the shortest noise wavelength was proposed, the system was tested in [3] with parameters relevant to cable length of 5% of the shortest noise wavelength.

**(ii)** Another tool to fight non-idealities is the cable capacitance compensation (capacitor killer) arrangement (see Figure 2). With practical cable parameters and their limits, there is a more serious threat of the security: the cable capacitance shortcuts part of the noise current and that results in a greater current at the side of the lower resistance end yielding an information leak. This effect can be avoided by a cable-capacitor-killer [3] using the inner wire of a coax cable as KLJN line while the outer shield of the cable is driven by the same voltage as the inner wire. However, this is done via a follower voltage amplifier with zero output impedance. The outer shield will then provide all the capacitive currents toward the ground and the inner wire will experience zero parasitic capacitance. Without capacitor killer arrangement and practical bare-wire line parameters, the recommended upper limit of cable length is much shorter depending on the driving resistor values $R_L$ and $R_H$. For example, the reported measurements in [3] used 5 kHz noise bandwidth and 2 km range for the capacitor killer arrangement. That was 5% of the wavelength corresponding to the 5 kHz cut-off frequency. However, without capacitor killer, the allowed upper limit for bare-wire arrangement would have been only 1% of the range limit mentioned above (see page 980 in [3]), that is, only 0.05% of the wavelength, which is 20 meters.

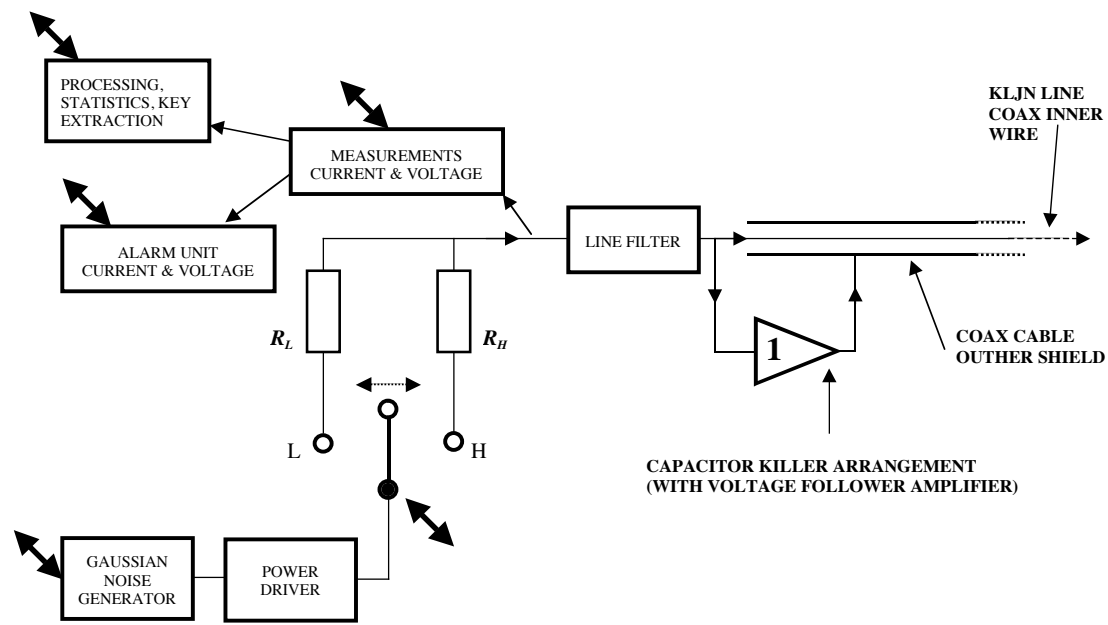

**Figure 2.** The tested KLJN cypher [3]. Double-ended arrows symbolize computer control.

## 2. Security: theory, circulators, cables and modeling.

In this section, we focus on the physical aspects and implications of the claims in [1].

### 2.1 The "no-wave" frequency regime

First, we provide the valid physical picture about cables in the no-wave limit. These considerations are well known for a very long time and we use standard textbook material, as well as well-established classical theories and pictures [13-15]. We first address the theory describing cable reflections in [1] and [7].

Electromagnetic waves are described as wave solutions of the Maxwell Equations [13], where half of the wave energy is carried by the electrical field and the other half by the magnetic field. Stationary wave solutions in a finite size cable are given by a Fourier series consisting of propagating forward and backward wave modes. The lowest frequency wave mode's wavelength is the double of the cable length. Since no lower frequency wave modes exist in a cable, lower frequency signal components do not propagate as a wave, but as a retarded (no-wave) potential or a near field around an antenna. Thus, such a propagating potential lacks of basic wave properties, such as interference and reflection.

Therefore, Equations 1-6 in [1] containing reflection and transmission coefficients, which are intended to describe the original system [2] and the new circulator-based setup [1], are *invalid* because no reflection can exist at the given conditions of no-wave regime: cable length is at most 5% of the shortest signal wavelength, the essential regime of security [2,3,8]. A simple example from optics clearly illustrates this situation: using a 500 nanometers visible light for communication, the maximal allowed length of the optical fiber would be 0.05*500= 25 nanometers. This means that only about 100 atoms would separate the outputs of Alice and Bob: they are virtually collocated!

### 2.2 Cables and their simulation in the no-wave limit

Cables in the high-frequency limit can be described by a distributed $RLC$ network (see Figure 3). In this picture, wave means a propagating voltage/current disturbance with half of the energy in the capacitors due the voltage of the wave and half in the inductors due to the current of the wave. For example, driving a cable at one end with a voltage generator in the no-wave (low-frequency) limit with no load at the other end means no current because the impedance of the capacitors is large and thus there is no current and related energy in the inductors. Hence, all the electrical energy is present only in the capacitors and therefore not in a wave mode. Time delay effects in such case are no-wave types of retarded potentials and thus, they neither get reflected nor do interfere like waves do.

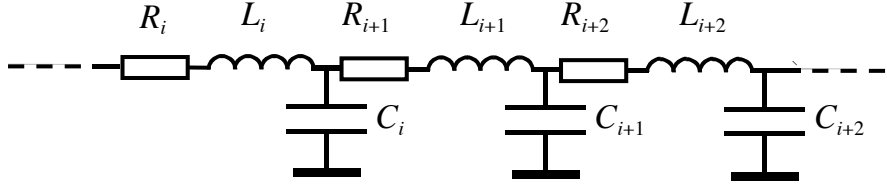

**Figure 3**. The cable is a distributed RLC network. Wave means a propagating disturbance where half of the energy is carried by the capacitors and the other half by the inductors. Other types of disturbances do not have wave properties, such as interference and reflection.

In the cable simulations presented in [1] the cables were characterized by their wave impedance $Z_0$. The wave impedance of the cable in the high-frequency limit is given as

$$Z_0 = \sqrt{\frac{L_0}{C_0}} \quad , \tag{7}$$

where $L_0$ and $C_0$ are the characteristic cable inductance and capacitance relevant for 1 meter cable length, respectively. It is important to recognize that the wave impedance alone does not determine the cable behavior at such short distances and provides only the ratio $L_0/C_0$, but not the actual values of $L_0$ and $C_0$. There is an *infinite* number of pairs of cable capacitance and inductance for a given wave impedance because only the ratio of $L$ and $C$ is determined. This has important implications. The resultant generator resistance driving the cable at the two ends will determine if the cable shows capacitive or inductive response. For example, in the case of high-resistance drive and short cable (no-wave regime), cable capacitance effects will dominate the transfer characteristics and the wave impedance does not contain enough information to determine the behavior because it leaves the actual capacitance value undetermined. Thus, for simulations in such a case like in [1], the wave impedance alone is irrelevant (see also Section 2.5).

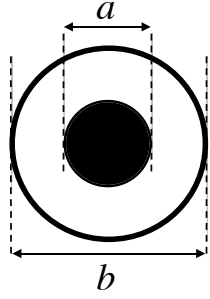

**Figure 4.** Cross-sectional view of the coaxial cable and the related definitions: *a* is the diameter of the inner wire and *b* is the internal diameter of the outer shield.

In this paper, we focus on coaxial (coax) cables (see Figure 4) because of their potential for capacitor killer arrangement. However sometimes we will also show that two-wire lines result in a similar conclusion. The approximate characteristic capacitance and inductance of coax cables are given as [14,15] :

$$C_0 = \frac{2\pi\varepsilon_r\varepsilon_0}{\ln(b/a)} \quad \text{(Farad/meter)} \quad , \tag{8}$$

$$L_0 = \frac{\mu_r\mu_0\ln(b/a)}{2\pi} \quad \text{(Henry/meter)} \quad . \tag{9}$$

With the practical values of the relative permeability $\mu_r = 1$, the wave impedance is [14,15]:

$$Z_0 = \frac{138}{\sqrt{\varepsilon_r}}\ln(b/a) \quad (\Omega/\text{meter}) \tag{10}$$

and, in case of given wave impedance, inner wire diameter *a*, and relative dielectric constant $\varepsilon_r$ , the required inner diameter of shield (the lower limit of coax cable diameter) is:

$$b = a\,\exp\frac{Z_0\sqrt{\varepsilon_r}}{138} \quad \text{(meter)} \tag{11}$$

Supposing a coax cable and $Z_0 = 100$ Ω (see Figure 4a in [1]), and dielectric constant $\varepsilon_r = 1$, $a=1$ mm, we obtain $d= 2.1$ mm. The characteristic inductance (Equation 9) is $L_0=1.48*10^{-07}$ Henry/meter and the total inductance of the simulated 2 kilometers line in [1] is $L= 2.96*10^{-4}$ Henry. The impedance of that inductance at the allowed highest frequency in the noise bandwidth is $Z_L(5kHz) = 9.29$Ω which is far below the value $R_w = 200$ Ω of practical resistance of the cable in [3]. Thus, the voltage drop in the wire is purely dominated by the resistance (Bergou-Scheuer-Yariv effect [3]).

The lower limit ($\varepsilon_r = 1$) of characteristic capacitance is $C_0=8*10^{-11}$ Farad/meter, which is at the lower end of practical values. Thus, the lower limit of total cable capacitance for the 2 kilometers length in [1] is $C=1.6*10^{-7}$ Farad.

To show the non-existence of reflections from another aspect, we give a pessimistic estimation about the possibility of oscillations by concentrating the distributed *RLC* elements of the cable into discrete circuit elements (see Figure 5). The quality factor *Q* of the concentrated *RLC* serial resonator is:

$$Q = \frac{L/R}{\sqrt{LC}} = 0.0228 \tag{12}$$

where *R* is now the sum of $R_w = 200$ Ω and the resultant diving resistance ($1.69 k\Omega = 2k\Omega \| 11k\Omega$) used by Alice and Bob.

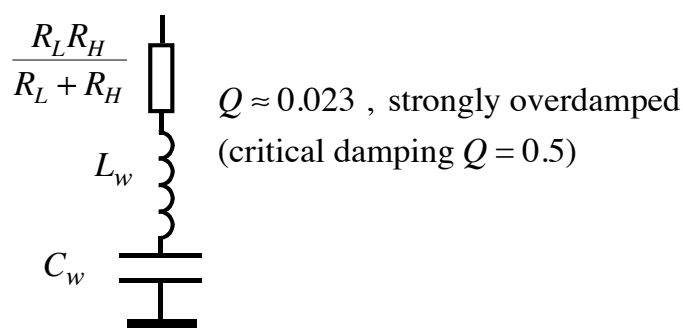

**Figure 5**. The cable capacitance, cable inductance, cable resistance and driving resistance can be concentrated into lumped circuit elements when the cable length is much shorter than the signal wavelength. When the driving resistance is much greater than the wave impedance of the cable, the cable acts as a capacitor and the inductance can be neglected.

Such an extremely low *Q* factor means that the cable with these parameters is a totally overdamped system with neither reflections nor oscillations because the critical limit of the overdamped (non-oscillating) range is $Q = 0.5$. Below this limit, the system shows no oscillation but a monotonic relaxation. This is another, phenomenologic interpretation why reflections cannot occur in the frequency range of the no-wave regime. This holds not only for the concentrated circuitry, but also for the cable with the distributed elements. The reason is that, for the distributed elements, the $\left(2\pi\sqrt{L_i C_i}\right)^{-1}$ eigenfrequencies are higher and thus they are more beyond the noise bandwidth than those of the concentrated elements. When the driving resistance is much greater than the wave impedance of the cable, like here, the cable acts as a capacitor and the inductance can be neglected. Therefore, if any oscillation had been observed during the simulations that would have been an artifact (e.g. errors caused by stray frequency components due to aliasing issues, etc.).

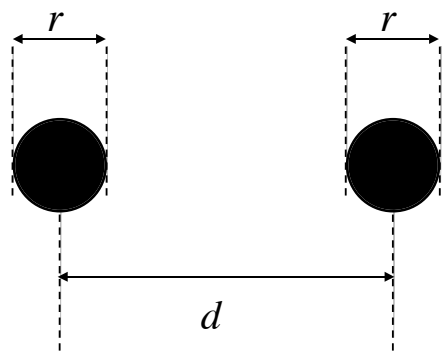

**Figure 6.** Cross-sectional view of the two-wire cable and the related definitions: $r$ is the diameter of the inner wire and $d$ is the distance between the centers of the wires ($d>r$).

For practical free-standing two-wire lines [14,15] we have the corresponding quantities (see Figure 6 for definitions):

$$C_0 = \frac{\pi \varepsilon_r \varepsilon_0}{\ln\left[(d - r_0)/r_0\right]} \quad \text{Farad/meter} \tag{13}$$

$$Z_0 = 276 \log_{10}(d/r) \quad, \quad \text{thus} \quad d = r\, 10^{Z_0/276} \quad, \tag{14}$$

with similar implications.

## 2.3 High-impedance simulation results in [1]: heavily unphysical parameters

Here we explore the implications of cable parameters used for the simulations in [1]. In Section 2.2, we have already pointed out that the wave impedance is a completely irrelevant parameter for the given combination of cable length and frequency band. We now show that the simulations used in [1] in the high wave impedance regime ($2\ k\Omega < Z_0 < 11\ k\Omega$) inherently assume heavily unphysical cable parameters. Below we suppose $\varepsilon_r=1$ and $\mu_r=1$ which are around the practical cable values. Using materials with $\varepsilon_r=30000$, the value around the maximum in ferroelectric ceramics, which are non-usable for cables, the predicted sizes below will shrink by a factor of $\sqrt{30000} \approx 170$ (see Equation 11) and they still stay unphysical or unpractical. We arrive at similar conclusions for high $\mu_r$ materials, too. Furthermore, using $\varepsilon_r=30000$, would further enhance the capacitance problems discussed below (in Section 2.4) and that 30000 times greater cable capacitance would kill the secure communication, even at 1% of the present range and at any wave impedance.

With a coax cable of $Z_0 = 11\ k\Omega$ (see [1]), inner wire diameter of 1 millimeter and $\varepsilon_r = 1$, according to Equation 11 in Section 2.2, the required coax shield diameter is $4.15*10^{31}$ meters. This is more than 28000 times greater than the size of the known universe (156 billion light years = $1.48*10^{27}$ meters [16])!

At $Z_0 = 4.7\ k\Omega$ (see the simulation data in Figure 4b in [1]), the required cable diameter is $6.18*10^{11}$ meters which is about the same size as the mean distance of the Earth and Jupiter ($7.79*10^{11}$ meters) [17].

At $Z_0 = 2\ k\Omega$ [1], which is the boundary case of the two observed type of behaviors of current and voltages in [1], the required cable diameter is 1.97 kilometers, which is basically the same size as the 2 kilometers range of communication, that is, the length of the cable in [1].

Though our focus is on coax cables, we note that the situation of two-wire lines produces even more unphysical values. With $r = 1$ millimeter and $Z_0 = 11\ k\Omega$, Equation 14 yields $d = 7.16*10^{36}$ meters which is more than *one billion times* greater than the known size of the universe (see above). With $Z_0 = 4.7\ k\Omega$ (Figure 4 b in [1]) $d = 1.1*10^{14}$ meters which is about 20 times greater than the planetaric-size of the solar system (the mean distance of the Sun and Pluto is $5.9*10^{12}$ meters [18]). With $Z_0 = 2\ k\Omega$, [1] the distance $d$ between the cables is 17.6 km, 8.8 times longer than the range of communication (2 kilometers), that is, the modeled cable length [1].

Therefore, we regard all the simulation results for the high impedance range, $2 \, k\Omega < Z_0 < 11 \, k\Omega$, including Figure 4 b in [1], as massively non-physical. They are unphysical because, among others, the cable equations and simulation programs are all based on the key assumptions that the *length of the cable is much greater than its diameter.*

It is an open question how the applied simulation program LTSpice accepted and processed the massively unphysical cable parameters. The clarification of this question is out of the scope of our paper. Though the unphysical nature of the large wave impedance parts of the simulations makes further considerations unnecessary, we also note that if the data in the LTSpice program did not get corrupted by the non-physical parameters then the observed difference of the voltages can be simply due to the cable inductance which would be huge at such diameters. When the cable inductance is large enough so that its impedance $Z_L(f) = 2\pi L$ becomes comparable to the resultant resistance of driving by Alice and Bob (see below), the inductance will separate Alice and Bob and a similar security breach takes place as with the wire resistance issue [4,7,8].

## *2.4 Low-cable-impedance simulation results in* [1]*: cable capacitance artifacts*

As mentioned above, according to earlier estimations (see [3], page 981) using the driving resistances discussed above, without capacitor killer arrangement only 1% of the nominal communication length (i.e. 0.05% of the wavelength) could have been utilized as range in [3]. Cable parameters in the simulations in [1] with the low wave impedances (see Figure 4a in [1]) violate this standard practical recommendation and assumption about handling the cable capacitance problems. The same high driving resistance values as in the experiments in [3] ($2k\Omega$, $11k\Omega$) are used in the simulations in [1] for the same range (2 kilometers), but without capacitor killer. Notice that this is 100 times longer cable than allowed. Thus the low wave-impedance simulation results are cable capacitance artifacts (see point (ii) in Section 1) and hence, in contrast to the conclusion in [1], they do *not* imply the insecurity of the KLJN cypher. Of course, not surprisingly, the KLJN cypher may become insecure in case of inappropriate use. Note, however, that remark 1 about unconditional security given in Section 1 still holds because Alice and Bob have full knowledge about Eve's information (see also Section 4).

With the simulated arrangements in [1], where no capacitor killer was applied and the length was at 2 km, the resistors of Alice and Bob were virtually shortcut by the cable capacitance in the high-frequency range, where the dominant portion of the noise power is located. The simple model of this case is shown in Figure 7, together with the relevant equations for the high-capacitance limit. The impedance of the capacitance is much less than $R_L$, therefore the capacitor represents a short circuit toward the ground and the currents at Alice and Bob's ends are simply the thermal noise currents of the noise voltage generators.

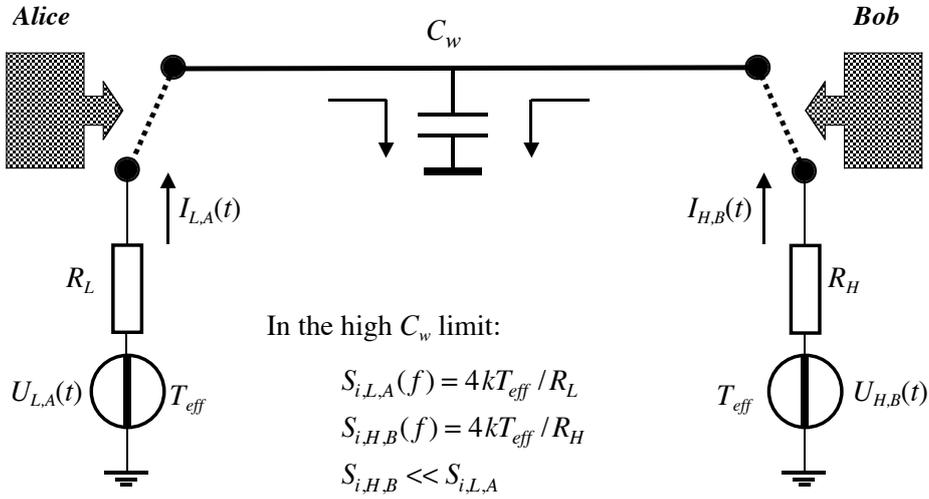

**Figure 7**. Illustration of the cable capacitance artifact leading to incorrect conclusion about the security of the KLJN cypher during the low wave impedance simulations in [1].

The current will be greater at the end with the lower resistance, just like it was observed in [1]. In the worst case, the ratio of the currents of the two sides is $R_H/R_L$. Supposing that Alice has the 2 $k\Omega$ and Bob the 11 $k\Omega$ resistance, the mean-square current at Alice's side can be 5.5 times greater than Bob's in the worst case. This implies that Bob's effective current amplitude is only 42% of the current at Alice's end. The current amplitudes shown in Figure 4a in [1] support this observation, even though we have no information about the value of the actual cable capacitance LTSpice used for the simulation.

Let us make a more accurate estimation with the practical cable data used in Section 2.2 by using the lower practical limit of total cable capacitance for the 2 kilometers length ($C=1.6*10^{-7}$ Farad). For the resultant driving resistance $R = 1.69\ k\Omega$ we get $f_c = 1/2\pi RC \approx 0.59\ kHz$. Since the total noise bandwidth is $5\ kHz$, the band $0.59\ kHz < f < 5\ kHz$ is cut off by the cable capacitance. This means that only about 12% of the total noise power is used for communication; the rest, 88% is shorted by the cable capacitance. This implies a serious security risk. Using higher dielectric constants would make this error progressively worse and, at the maximal value of $\varepsilon_r=30000$ mentioned above, no security would exist even at 1% of the 2 kilometers range.

On the other hand, using the usual $\varepsilon_r \approx 1$, at the 1% of the 2 km length (20 meters) would result in a cutoff frequency of 59.63 kHz and guarantee that the capacitive shunt currents stay sufficiently low to achieve the aimed raw bit information leak in the order of 1% [3, 8,10]. Similarly, using a capacitor killer arrangement designed to compensate out 99% of the cable capacitance at 2 kilometer distance would result in the same satisfactory situation.

Finally, note that Eve could use much more efficient tools than in [1], e.g. by calculating the mean product (crosscorrelation) of the time-derivative of the line voltage and the currents at the two ends of the line. The side where this product is greater has the *L* bit.

**3. Analyzing and generalizing the circulator-based key exchange proposed in [1]**

In [1] a new secure key exchange system utilizing circulators to emulate the KLJN cypher was introduced. Due to the lack of a detailed circuit analysis in [1] the security aspects and weaknesses of the circulator-based system were not clear. In this section we show a generalized circulator circuitry, deduce its relevant equations and show how to design such a secure communicator. Surprisingly, the knowledge of the Johnson noise equations is unnecessary for the security design. We show a major weakness of all circulator-based key exchange systems: *their absolute vulnerability against a circulator-based man-in-the-middle attack*. Finally, we point out that the original, circulator-free, Johnson noise based KLJN system is secure even against circulator based man-in-the-middle attacks.

## 3.1 The circulator circuitry and how does it work

Circulators are usually passive circuit elements for high-frequency applications. In [1], an active, circuit-based system is cited (reference [19] here) which emulates most of the essential properties of a circulator. We have analyzed the function of the circuitry in [19] and we constructed a *general* circulator circuitry (see Figure 8). Despite its simplicity with respect to the system in [19], it has the same function. Another advantage of the circuit in Figure 8 is that it does not change the sign of the voltage between ports, which is an artifact in [19].

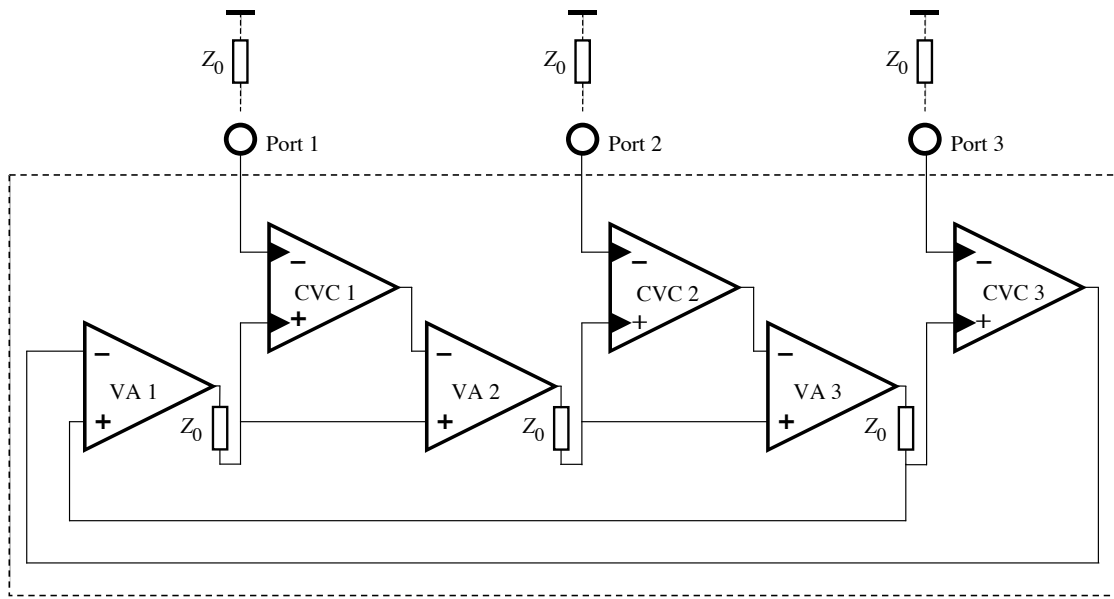

**Figure 8.** Generalized circuit scheme to emulate passive circulators (within the dashed line area). An improved version of the circuitry cited in [1] (Ref [19] here). VA stands for differential voltage amplifier with amplification of 2 and CVC stands current-voltage converter with amplification $Z_0$, respectively.

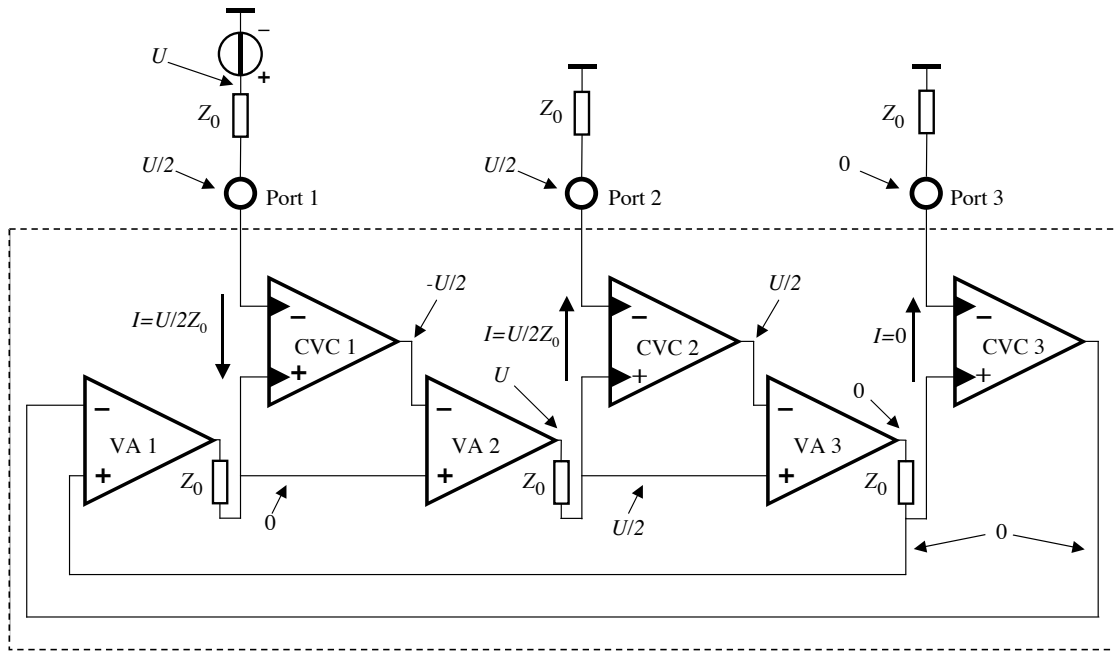

**Figure 9.** Explanation of the function of the general circulator circuit. Port 1 is driven by a voltage generator of voltage $U$ and internal resistance $Z_0$. Then a voltage of $U/2$ will drop on Port 1 and driven by the current $I$ flowing into Port 1, the circulator distills the same $U/2$ voltage on the next port (Port 2). The subsequent port (Port 3) stays at zero voltage because the current and voltage signals driving Port 2 will compensate to zero.

The function of the circulator in Figures 8 is easy to follow see Figure 9. A current induced through a port by an external voltage will yield a voltage at the next port, however, an internal voltage at the output of a VA will not result in any voltage at the next port, see the explanation below. The input/output resistance of the ports is $Z_0$ and they should be driven by the same external resistance value. The external resistances are grounded in this figure representing zero voltage drive. When one of them (Port 1 in Figure 9) is not grounded, but a voltage $U$ drives it via the external $Z_0$, a voltage of $U/2$ drops on that port. The same voltage ($U/2$) will be distilled on the next port, too. On the other hand, the corresponding current flowing out at the next port via the related CVC yields an internal voltage component at the output of that CVC which kills further voltage propagation at the input of the VA driving the subsequent (third port). Thus the third port stays at zero voltage. Because the system is linear, the ports can be run independently and the distillation will satisfy the linear superposition theorem. However, it is essential to recognize that these idealistic working conditions of the circulator can never be reached. Depending on the practical inaccuracies of the resistance values and amplification, the same order of relative errors will occur in signals on the relevant output ports. Furthermore stray signals with the same order of relative amplitude will occur on ports which should show zero voltage at ideal conditions. Thus, in the secure key exchange schemes analyzed below, practical inaccuracies in the circulator will lead to additional information leak, out of the known ones, such as the non-zero cable resistance.

## 3.2 Generalized circulator schemes (unrelated to Johnson noise) for secure key exchange

In Figure 10, the generalization of Liu's circulator-based communicator (Liu-cypher) is shown. The Liu-cypher works by emulating the Johnson noise situation, that is, the net power flow between Alive and Bob is zero. It is unnecessary to make Johnson noise calculations because the system can securely designed without the knowledge of the corresponding Johnson noise situation. We will point out that the best practical situation is when only one side is using a circulator, the other one is using just a resistance $Z_0$ and a noise voltage generator.

The generalized circulator-cypher in Figure 10 consists of a noise generator at each side ($U_A(t)$ and $U_B(t)$, respectively), a circulator, and a voltage amplifier. The voltage amplifier outputs, with distilled voltage originating from the other end of the line, are added to the noise generator voltages. The resultant channel voltage and current contributions originating from Alice are given by

$$U_{ch,A} = U_{A,tot}(1+B)/2 \quad \text{and} \quad I_{ch,A} = U_{A,tot}(1-B)/2Z_0 \quad , \qquad (15)$$

respectively. The resultant channel voltage and current contributions from Bob are:

$$U_{ch,B} = U_{B,tot}(1+A)/2 \quad \text{and} \quad I_{ch,B} = U_{B,tot}(1-A)/2Z_0 \quad . \qquad (16)$$

With proper distillation factors $A$, $B$ and noise generator voltages $U_A(t)$, $U_B(t)$, the security is ensured by running the system with zero net power flow. The security condition, that is, the requirements for zero average power flow between Alice and Bob, are as follows (the linear superposition theorem and the statistical independence of the noise generators of Alice and Bob are supposed):

$$\langle U_{ch,A}\ I_{ch,A}\rangle_t = \langle U_{ch,B}\ I_{ch,B}\rangle_t \quad \Rightarrow \quad \langle U_{A,tot}^2(t)\rangle(1-B)^2 = \langle U_{B,tot}^2(t)\rangle(1-A)^2 \quad (17)$$

One might argue that the line-node resistors of the circulators of Alice and Bob dissipate equal energy, however this is irrelevant because that neglects the fact that the real issue is the power balance between the correlated voltage generators at the two ends. The line-node resistors have always the same current, because this is a single loop. Therefore, their power dissipations are always trivially equal, even when their total power balances are different.

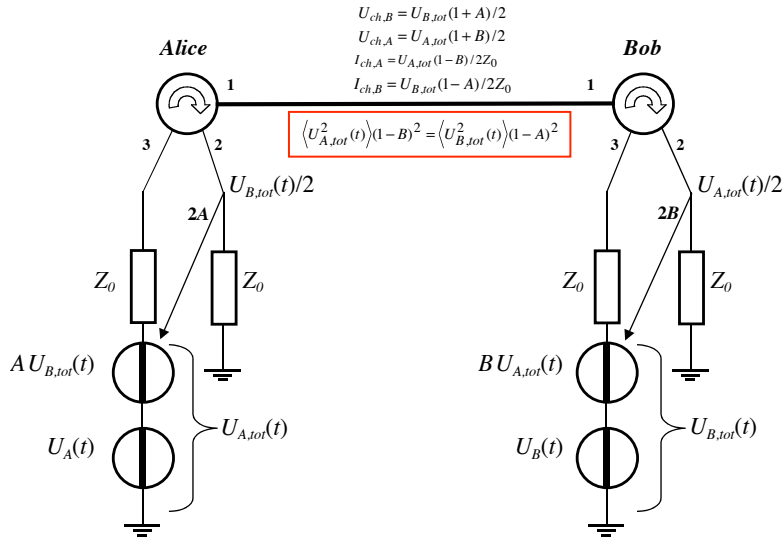

**Figure 10.** The generalized circulator-based secure key exchange to replace the Liu-cypher proposed in [1]. Voltage amplifier outputs with distilled voltage of the other end are added to the noise generator voltages. With proper distillation factors $A$, $B$ and noise generator voltages $U_A(t)$ and $U_B(t)$, the system can run with the zero net power flow condition to maintain security. The Johnson noise equations are not needed for the design. In the best circulator-based system either $A$ or $B$ is zero, which means that one of the circulators and amplifiers can be left out, see Figure 11.

To determine the needed noise voltage generator contributions by Alice and Bob, we must realize that the system has a *positive feedback loop* at nonzero $A$ and $B$ because the voltage generated at one side will be distilled at the other side and fed into the line there. However, that distilled voltage will also be re-distilled and added to the original voltage at the first side; and so on. Positive feedback amplifies the incoming noise voltages $U_A(t)$ and $U_B(t)$ and the total voltage contributions of the two sites are

given as:

$$U_{A,tot}(t) = \frac{U_A(t)}{1-AB} \quad \text{and} \quad U_{B,tot}(t) = \frac{U_B(t)}{1-AB} \quad . \quad (18)$$

Equation 18 is deduced from

$$U_{A,tot}(t) = U_A(t) + AU_{B,tot}(t) \quad \text{and} \quad U_{B,tot}(t) = U_B(t) + BU_{A,tot}(t) \quad (19)$$

by considering one of the cases of $U_A(t) = 0$ or $U_B(t) = 0$. It can be seen from Equations 19 that the positive feedback's loop-amplification ($AB$) will enhance the input noise generator voltages of Alice and Bob by the same factor.

After these preparations, the design of a secure circulator-based system is straightforward by using Equations 14-18 and the knowledge and application of the *Johnson noise equations* (current and voltage [2]) *are unnecessary* unless the task is to exactly emulate a concrete situation with Johnson noise.

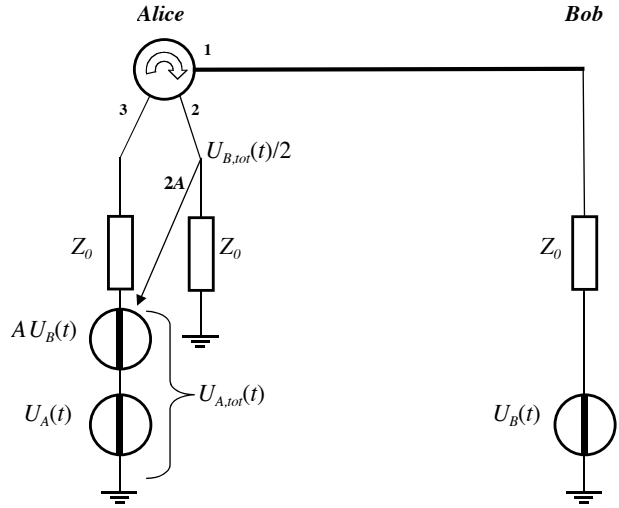

**Figure 11**. The simplest circulator-based cypher which is also the most robust because there is no positive feedback loop in this system therefore the nonideality, bandwidth and transient problems are kept at the lowest level here.

Though the conceptual/idealized Liu-cypher (Figure 10) is secure against passive attacks (but not against the man-in-the-middle attack, see Section 3.3) the positive feedback loop in the circulator system with $A > 0$ and $B > 0$ coefficients raises serious concerns about the practical security. It follows from the well-known facts in basic electronics that any non-idealities, such as nonlinearity, frequency-dependence, wire resistance, wire capacitance, will be enhanced by a positive feedback loop. This situation will help Eve to enhance the information leak normally present in any practical KLJN cypher. Moreover, it opens new information leak channels due to the circulator circuit non-idealities. Therefore, these effects must be thoroughly analyzed and addressed during an actual design and the tools of necessary protection installed.

However, this situation can be improved by avoiding a positive feedback loop with circulator only at one end. The modified circulator-based cypher is shown in Figure 11. In the figure, only Alice has circulator; Bob's side has only a noise generator and $Z_0$. To enhance security the decision of using a circulator or not should be made randomly at each clock period. Clearly, only the asymmetric situations of Alice and Bob count as secure bit.

### *3.3 Cracking all the circulator-based key exchange schemes*

We now show that Liu-cypher and its generalizations discussed above are vulnerable against even a simple attack utilizing circulators. It is important to note that the attack described below does not address small information leaks like the ones generally present in any practical (classical and quantum) communicators (see Sections 1 and 2) and which can be removed by privacy amplification. The issue here is fatal. We show that Eve can extract exactly the same amount of information as Bob and Alice share. That is, no secrecy remains implying that no privacy amplifier or other method can be used to guarantee security. The major conceptual disadvantage of any of the circulator-based cyphers (including the Liu-cypher) is that they are absolutely vulnerable to man-in-the-middle attacks executed by similar circulators (see Figure 12). This is because Eve can break the line and install a properly wired circulator pair, which is basically a circuit acting as a *repeater* in both directions. Node-1 of circulator Eve-A is connected to Alice via the line and Eve-A distills half of Alice's total voltage ($U_{A,tot}/2$) at its node-2. That node, after a voltage amplification of a factor of 2, will drive node-3 of Eve's other circulator, i.e. Eve-B. Thus, idealistically, Node-1 of Eve-B will exactly emulate the original voltage ($U_{A,tot}$) and current contribution of Alice toward Bob. The same holds in the reverse direction, too. Node-1 of circulator Eve-B is connected to Bob via the line and Eve-B distills half of Bob's total voltage ($U_{B,tot}/2$) at its node-2. That node, after a voltage amplification of a factor of 2, will drive node-3 of Eve's other circulator, Eve-A. Thus, idealistically, Node-1 of Eve-A will exactly emulate the original voltage ($U_{B,tot}$) and current contribution of Bob toward Alice. In this way, the feedback loop is closed and the original functions of the circulators of Alice and Bob are restored by Eve's repeater. However Eve can measure $U_{A,tot}$ and $U_{B,tot}$ therefore she identifies the shared bit with the same accuracy as Alice and Bob; while Eve extracts the whole information, the current and voltage based alarm system does not recognize the attack situation.

In conclusion, the circulator-based key exchange protocols have zero protection against a circulator-based man-in-the-middle attack, as stated.

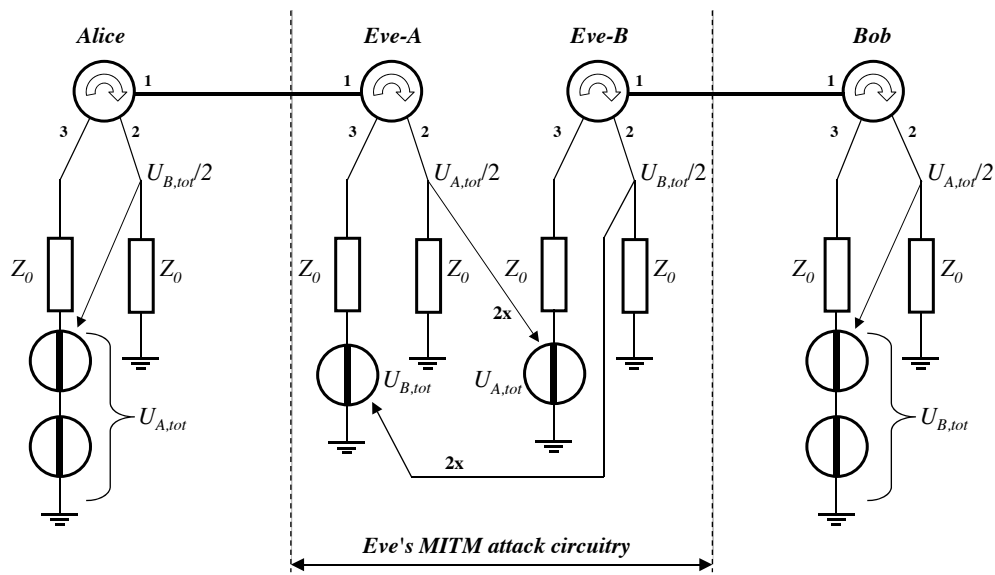

**Figure 12**. Cracking all circulator-based communicator schemes (including the Liu-cipher, the generalized cipher and the cipher on Figure 10) by a circulator-pair based man-in-the-middle attack.

## *3.4 Robustness of the security of the original Johnson noise based scheme*

It can be shown that, using two *different* pairs of properly modified (asymmetric) circulator circuits, or two different wiring of an asymmetric circulator pair, corresponding to situations of the different resistance values used by Alice and Bob, a modified circulator-based man-in-the-middle attack can also be carried out against the original KLJN cypher. Asymmetric circulators, circulators with different node resistances, are needed for making a repeater because of the asymmetric resistor values of Alice and Bob during exchanging a secure bit. However, because Eve does not know which resistor Alice and Bob will choose, Eve does not know which arrangement of the circulators should be used. Her chance to succeed with a single bit is 25% because there are 4 distinct possibilities. That is, 75% is the probability that there is a mismatch and current/voltage based alarm will go on during a single bit extraction attempt by Eve. Thus the probability that Eve can stay hidden for $N$ bits is $(0.25)^N$. Note, only half of those bits will be useful because the *LL* and *HH* cases can't be used. Thus Eve's chance that she can extract 10 key bits without igniting the alarm is less than one trillionth. In conclusion, the original Johnson-like noise based KLJN system is secure even against a circulator-based man-in-the-middle attack.

## 4. Key Exchange by Dropping High-Risk Bits

We now comment a second remark in [1] stating: Alice and Bob "*can drop high risk bits from the stream to decrease the risk [2]. Such an approach will reduce the effective key generation rate and is subject to further crypto-analysis.*"

We first note that the high-risk bits destined to be dropped in [2] are restricted only to the *LL* and *HH* bits because the state of the system is always publicly known for these types of bits. Accordingly, the locations of *all* such bits are also publicly known and hence, these bits can be dropped without loss. Clearly, this transformation will halve the key generation rate, as the *LL* and *HH* bits form about 50% of all the bits. We also note that dropping the *LL* and *HH* bits is done automatically at the very beginning and is a trivial but essential step of the secure key exchange method described in [2]. Neither Alice and Bob nor Eve will try to use these bits anymore. Thus, in what follows, we focus only on sequences not containing any *LL* and *HH* bits. Such sequences will be referred to as *basic* sequences.

Besides *LL* and *HH* bits, we have to consider the other type of high-risk bits causing information leak due to non-idealities, stray components, inaccuracies, etc. This type of high-risk bits is of completely different nature (see [1] and [2] for a detailed discussion). To comment the remark cited above from [1] for these types of high-risk bits, we first recall that Alice and Bob have full knowledge of the system state, as well as of the whole information (current/voltage data) Eve can have. To the best of our knowledge, this is a unique characteristic feature of the *KLJN* cypher among the existing physical secure key exchange systems that can be utilized in improving the security. Moreover, Alice and Bob also know the best strategy for the most efficient bit estimation. Since they can assume without loss of generality that Eve uses this best strategy, Alice and Bob also know her guess, or equivalently, they know whether or not Eve's guess is correct. On the other hand, Eve knows the probability $P$ ($>0.5$) by which she can successfully guess the *LH* and *HL* bits in the underlying system. However, she never knows if her guess of a particular *LH* or *HL* bit is correct or not. We recall that the success probability $P$ can be derived from the system parameters; we consider the worst-case scenario that these parameters are publicly known. Below we assume the much worse case that Eve knows not only $P$, but even the number $M$ of correctly guessed *LH* and *HL* bits. This assumption enables a simplified analysis of the number of potential keys. For the rest of this section, by high-risk bits we will mean the *LH* and *HL* bits correctly guessed by Eve. Furthermore, we will always assume that the underlying key exchange protocol is also known for Eve. Before going into the details, we note that if Alice and Bob use the entire basic sequence as final key then, for all basic sequences, the number of potential keys for Eve is exactly $\binom{N}{M}$. This number will be regarded as the *baseline* measure for the risk.

a) Consider first the key exchange protocol that Alice and Bob drop *all* high-risk bits from the basic sequence. Since Eve knows $N$ and $M$, as well as the protocol, she can obtain the set of potential keys for this case by dropping $M$ bits from the basic sequence in all possible ways. Thus, $\binom{N}{M}$ remains an upper bound on the number of potential keys and hence, this protocol does not decrease the baseline risk. It can, however, increase it because the sharpness of the above bound depends on the basic key eavesdropped by Eve. As an extreme example, consider the case that Eve's basic sequence consists of only ones. Clearly, she can easily reason that the final key is the sequence consisting of $N-M$ zeros. Thus, dropping *all* high-risk bits may only *increase* the baseline risk.

b) We now show that there exists no *deterministic* strategy based on dropping any subsets of the high-risk bits from the basic sequence such that it results in an increase of the security (i.e, in a decrease in the baseline risk). In fact, any such deterministic strategies may result only in a decrease in the security (see the case above for an extreme situation). More precisely, suppose that Alice and Bob determine the final key by a function $T$ assigning a sequence $k$ to the pair $(k_{AB}, k_E)$, where

  a. $k_{AB}$ is the basic sequence communicated by Alice and Bob,
  b. $k_E$ is the sequence eavesdropped by Eve, and
  c. $k$ is obtained from $k_{AB}$ by dropping some high-risk bits.

For this protocol, Eve can guess $k_{AB}$ from her sequence $k_E$ in $\binom{N}{M}$ different ways and then, as she knows $T$ by the worst-case assumption, apply $T$ to $(k'_{AB}, k_E)$ for every guess $k'_{AB}$. The number of potential keys is therefore also at most $\binom{N}{M}$, i.e., even this more general strategy based on dropping any subsets of the high-risk bits does not decrease the risk. Notice, however, that the above bound is sharp only when $T$ is injective in its first argument; all other choices of $T$ do increase the baseline risk. We also note that the argument above holds for any arbitrary function $T$, i.e., $T$ need not be restricted to dropping high-risk bits.

c) Finally, we consider the case that Alice and Bob do not fix some deterministic strategy in advance and generate the final key from the basic sequence by (i) selecting first the high-risk bits to be dropped with some unknown strategy and (ii) exchanging then the positions of the bits kept. As an example of this protocol, consider the case that Alice first selects the high-risk bits to be dropped by some stochastic method (e.g., simple coin toss) and sends then these bit positions to Bob. (For the authentication, she can again utilize the KLJN cypher, as it is robust against the "man-in-the-middle" attack [12].) We regard again the worst-case that Eve is aware of the number $M'$ ($M' < M$) of high-risk bits dropped from the basic sequence, as well as of the $N$-$M'$ bit positions kept from the basic sequence. Using these assumptions, for this case she can determine the set of potential keys by

  1) dropping first the same $M'$ bits from her eavesdropped sequence,
  2) choosing $N$-$M$+$M'$ positions in the sequence obtained, and

3) inverting each bit in all selected positions for every choice.

Since different choices in step 2) result in different keys in step 3), the number of potential keys is equal to the number $\binom{N-M'}{N-M+M'} = \binom{N-M'}{M-M'}$ of possible choices. As $\binom{N}{M} > \binom{N-M'}{M-M'}$ always holds for every $M' > 0$, dropping only high-risk bits increases the risk for this protocol as well and is therefore a non-optimal strategy.

A more generalized and practical crypto analysis of the privacy amplification of the non-idealistic KLJN cypher, including the situation when Eve works with the combination of the various strategies and utilizing system parameter fluctuations, is currently in progress by the present authors.

**Conclusions**

After analyzing the relevant cable and circuit systems, we have refuted most of the claims in [1] concerning the security of the circulator-based key exchange system and the theory of the KLJN cypher. In particular, in contrast to the conclusion of [1], we have shown that:

**(i)** the circulator-based key exchange system described in [1] is insecure and

**(ii)** the arguments in [1] do not break the security of the KLJN system [2].

We have also studied a remark in [1] about a security-enhancing method based on dropping high-risk bits and shown for different protocols that this method does not increase the security. Our unpublished preliminary results however indicate that this method could still be useful if not only high-risk bits are dropped from the sequence communicated by Alice and Bob.

**Acknowledgements**

(December 2008) was supported by a TAMU travel grant for sabbatical leave.